\documentstyle[twoside,fleqn,espcrc2,epsf]{article}

\newcommand{\AmS}{{\protect\the\textfont2
  A\kern-.1667em\lower.5ex\hbox{M}\kern-.125emS}}

\def\plotone#1{\centering \leavevmode
\epsfxsize=0.45 \textwidth \epsfbox{#1}}

\title{The X--ray spectrum of the polar BY Cam}

\author{C. Done \address{Department of Physics, University of Durham,
        South Road, Durham DH1 3LE, UK }
        and 
        P. Magdziarz\address{Astronomical Observatory, Jagiellonian University,
        Orla~171 30-244 Cracow, Poland}}
       
\begin{document}

\begin{abstract}
ASCA and GINGA X-ray data from the magnetic Cataclysmic Variable BY Cam show
that the spectrum is strongly affected by complex absorption, probably from the
pre-shock accretion column.  The intrinsic emission from the shock is
significantly better described by the theoretically expected multi-temperature
structure rather than a single temperature plasma, but with cyclotron cooling
probably suppressing the highest temperature bremsstrahlung components.  Reflection of
this multi-temperature emission from the white dwarf surface is also
significantly detected. All these spectral complexities are required to gain a
physically self-consistent picture of the X-ray spectrum. 

\end{abstract}

\maketitle

\section{INTRODUCTION}
Magnetic cataclysmic variables (polars or AM Her stars) are binary systems where
a magnetised ($B\sim 10^7$ G) white dwarf accretes from a low mass companion
(see the upper panel of Figure 1 and e.g. the review by \cite{Crop90}). Such magnetic
fields are strong enough to disrupt disk formation. Instead, the accreting
stream is entrained by the magnetic field and falls freely through the
gravitational potential until it hits the white dwarf surface. The resultant
strong shock has a typical temperature of tens of keV for optically thin
material, giving rise to an X--ray emitting plasma. 

\begin{figure}[htb]
\plotone{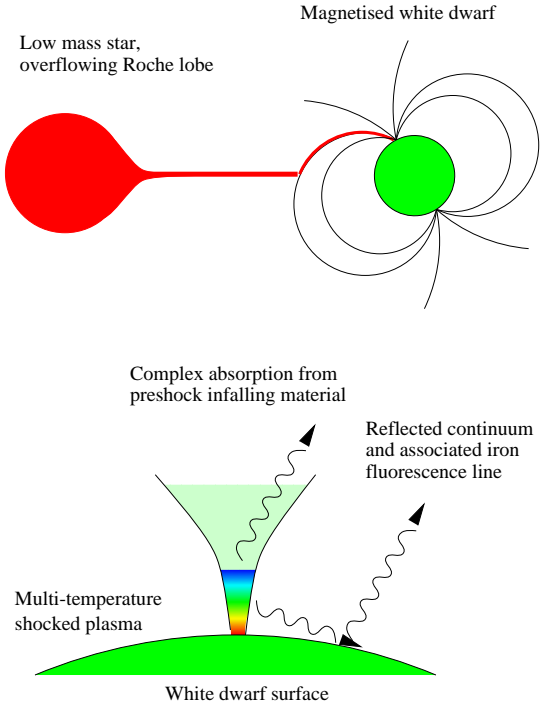}
\vskip -20pt
\caption{The geometry of polar systems}
\end{figure}

\section{SPECTRAL COMPLEXITY}

The shock clearly must have temperature structure as the heated post-shock gas
must cool in order to settle onto the white dwarf surface. Thus the primary hard
X-ray spectrum of a polar is expected to be the density weighted sum of
optically thin spectra with temperatures ranging from the shock temperature to
effectively the photospheric temperature of the white dwarf surface (see the
lower panel of Figure 1 and e.g. \cite{ID83,Done95}).  The
spectral signature of this `cooling flow' is that there is both high and low
temperature gas present. Thus high temperature bremsstrahlung continuum and high
ionisation iron K lines can be seen together with low ionisation iron L
lines. Figure 2 shows this difference between the expected emission from a shock
of maximum temperature of 25 keV, compared to a single temperature 10 keV
plasma.

The primary emission illuminates the white dwarf surface and has some
probability of being reflected (see lower panel of figure 1). The reflection
probability is given by a trade--off between the importance of electron
scattering and photo--electric absorption.  Since the latter is energy
dependent, the albedo is also energy dependent, with higher energy photons being
preferentially reflected due to the smaller photo--electric opacity of the
material. This gives rise to a reflected continuum spectrum that is harder than
the incident spectrum in the 2--20 keV GINGA range, with an iron K$\alpha$
fluorescence line superimposed on it. Figure 2 shows the reflected spectrum
expected from the shock spectrum illuminating the cool white dwarf surface.

The X--ray shock spectrum can also be distorted by absorption from the pre--shock
infalling material. This material is irradiated by the X--ray emission, so
should have a complex ionization structure (see e.g. \cite{RF80,Swank84,
Kall93,Kall96}). It is also spatially extended over the
X--ray source so that different segments of the X--ray emission travel through
different path lengths of the material \cite{Done95}, especially as the
accretion column is probably arc--like in cross-section, rather than circular
(e.g. \cite{Crop90}). The column should also add further complexity from
secondary emission \cite{Kall93,Kall96} and scattering \cite{Ish97}. 
Modelling including all these are being developed \cite{Rai97},
but here we use a phenomenological description of this complex
absorption in terms of a continuous (power law) distribution of neutral
absorping column and covering fraction ($C_f\propto N_H^\beta$, see also \cite{NWK91}). 

\begin{figure}[htb]
\plotone{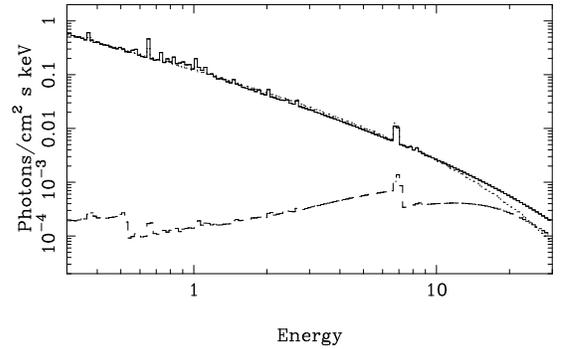}
\vskip -20pt
\caption{A 25 keV multi--temperature 'cooling flow' plasma has both iron K lines at
6.7--6.9 keV and iron L lines at 1 keV (solid line), while a single temperature 10 keV plasma which
has a rather similar spectral shape has only the iron K lines (dashed line). The expected
reflected spectrum from the multi--temperature plasma emission is also shown
(dotted line).}
\vskip -20pt
\end{figure}

\section{THE ASCA AND GINGA SPECTRA OF BY CAM}
Figure 3 shows the ASCA spectrum of BY Cam, a bright polar.
The spectrum is very hard, and fitting it by a single temperature plasma gives a
very high temperature of $kT=200_{-35}^{+170}$ keV ($\chi^2_\nu=1458/1272$).
The expected multi--temperature plasma models do not help, since they include
the softer cooling components, so give $kT_{max}=940^{+\infty}_{-310}$ keV
($\chi^2_\nu=1455/1272$). Clearly there is some distortion present that is
hardening the observed spectrum in the ASCA bandpass. Absorption is the most
obvious way to do this since Compton reflection only contributes significantly
to the spectrum above 5 keV, so is unlikely to strongly affect the ASCA data.
Neither simple absorption, nor partial covering, nor ionised absorption result
in a smooth (rather than abrupt) hardening of the spectrum, so we use a power
law distribution of column with covering fraction to approximate the complex physics.
This gives a significantly better fit to the spectrum, with a
physically resonable derived maximum temperature of $kT_{max}=45_{-14}^{+35}$
keV for a maximum column of $2.5^{+\infty}_{-1.8}\times 10^{24}$ cm$^{-2}$, with
index $\beta =-1.07\pm 0.08$.

While the resultant fit is statistically adequate, there are clear residuals
left around the iron K line. Adding a narrow Gaussian line at 6.4 keV and
reflection continuum that should accompany it gives $\chi^2_\nu=1360/1268$, with
a line equivalent width of $75\pm 25$ eV and a reflection continuum
normalisation of $f=1.5_{-1.3}^{+0.6}$ (where $f=1$ denotes the normalisation
expected from an isotropically illuminated slab covering a solid angle of
$2\pi$) and $kT_{max}=38^{+34}_{-13}$ keV.  Both line and continuum are
significantly detected, and the level of both is consistent with that expected
from the geometry.

\begin{figure}[!htb]
\vskip -30pt
\plotone{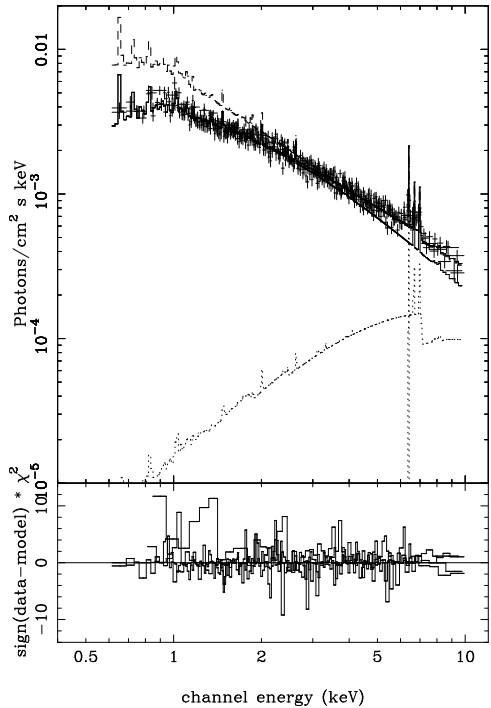}
\vskip -20pt
\caption{The ASCA spectrum of BY Cam, showing the intrinsic multi--temperature
emission (dashed line) and its reflection spectrum (dotted line). The solid line
shows the resultant spectum after complex absorption from the pre--shock
column. The bottom panels show the residuals of this fit, indicating that it is
a good description of the data.}
\end{figure}

Substituting the primary continuum for a single temperature plasma model (and
its reflection and including complex absorption as above) gives a much worse fit
with $\chi^2_\nu= 1380/1268$ for $kT = 18_{-6}^{+5}$ keV. The data contain
significant iron L line emission which cannot be fit by the single temperature
models.  This is the first observational confirmation of the theoretically
expected cooling of the shocked plasma in polars.

There are also GINGA data on BY Cam. These spectra extend from 2--20 keV so give
better constraints on the high energy spectrum. Fitting the same models as above
gives a maximum temperature of the cooling plasma of $kT_{max}=21^{+18}_{-4}$
keV.  Assuming that the shock only cools via bremsstrahlung then this 
implies that the mass of the white dwarf is $0.6^{+0.3}_{-0.1} M_\odot$
(see e.g. \cite{FKR92}). While the mass of the white dwarf in BY Cam
is not well constrained observationally, it is known that it has a substantial
magnetic field of 28 MG \cite{Sch96} so cyclotron cooling should also be
important. Cyclotron cooling can dominate over bremsstrahlung for
high temperature, low density material (see e.g. \cite{Wu95}), so it has the
effect of reducing the maximum observed X--ray bremsstrahlung temperature. 
For reasonable mass accretion rates it seems quite likely that the shock in BY
Cam has a composite structure, cooling via both cyclotron and bremsstrahlung
emission.

\end{document}